\documentclass[%
 aip,
 amsmath,amssymb,
preprint,%
]{revtex4-1}

\usepackage{graphicx}
\usepackage{dcolumn}
\usepackage{bm}
\usepackage[mathlines]{lineno}
\usepackage[utf8]{inputenc}
\usepackage[T1]{fontenc}
\usepackage{mathptmx}
\usepackage{etoolbox}
\usepackage{hyperref}
\usepackage{cleveref}

\makeatletter
\def\@email#1#2{%
 \endgroup
 \patchcmd{\titleblock@produce}
  {\frontmatter@RRAPformat}
  {\frontmatter@RRAPformat{\produce@RRAP{*#1\href{mailto:#2}{#2}}}\frontmatter@RRAPformat}
  {}{}
}%
\makeatother
\begin{document}

\preprint{AIP/123-QED}

\title[]{Splashing-regime transitions and secondary-droplet scaling in oblique drop impacts on a deep pool}
\author{Ying Xiong}
\author{Guangxi Cui}
\author{Yang Zhang}%
 \email{zhangyang@gdou.edu.cn}
\author{Xiaolei Li}
\author{Lingling Xie}
\affiliation{ 
Department of Ocean and Meteorology, Guangdong Ocean University, Zhanjiang, Guangdong, China.
}%

\date{\today}

\begin{abstract}
Oblique drop impact onto a deep liquid pool produces asymmetric crowns, directional jetting, and splashing transitions that cannot be characterized by the total impact inertia alone. We numerically investigate water drops impacting a quiescent deep pool over $41\leq We\leq1790$ and $10^\circ\leq\theta\leq90^\circ$. The simulations reproduce the principal features observed experimentally and identify five post-impact regimes in the $We$--$\theta$ plane: deposition, front splashing, side splashing, side-front splashing, and crown splashing. The deposition--front-splashing transition is described by the tangential-inertial parameter $K_s=We\cos\theta$, with $K_s^c\approx120$. This criterion follows from the competition between downstream crown-rim inertia and capillary retraction at the Taylor--Culick velocity. The transition from front to side-front splashing is instead governed primarily by normal impact inertia, with a critical normal Weber number $We_N^c\approx318$. Beyond these regime transitions, the secondary-droplet statistics reveal fragmentation behavior common to the different splashing regimes. The droplet-size distributions are positively skewed, and the median diameter follows $d_{s,\mathrm{med}}/D\sim We^{-3/5}$. Second-order longitudinal velocity structure functions are used to estimate the root-mean-square longitudinal velocity increment at the median-droplet scale. Combining this estimate with a capillary--inertial breakup condition recovers the observed diameter scaling. A closure based on the weak $We$ dependence of the normalized secondary-droplet volume and size-distribution shape then yields $N_s\sim We^{9/5}$, providing a scaling-based interpretation of the secondary-droplet-number dependence observed experimentally.
\end{abstract}

\maketitle

%

\section{\label{sec:level1} Introduction}
Raindrop impacts on the ocean surface affect numerous microscale air--sea interaction processes. The associated momentum transfer can facilitate the dispersion of surface oil slicks \cite{Li98relation,Thorpe95vertical,Murphy2015}, enhance near-surface turbulent mixing \cite{Lang2000droponmix}, and modify sea surface roughness by generating transient features such as cavities, crowns, ring waves, and Worthington jets \cite{Liu2018JPO}. These microscale structures can, in turn, affect radar backscattering and influence the remote sensing of rain cells \cite{Liu2016JGR}. However, direct field observation of raindrop impact under realistic conditions remains challenging.
Laboratory studies of single drops impacting deep pools have been conducted for over a century, beginning with the pioneering work of Worthington \cite{Worthington1908} and continuing through extensive research \cite{Edgerton1954,Jayaratne64,Ching84,Rodriguez85,Hsiao88,Rein93rev,Cresswell95,Rein96,Liow01,Leneweit05oblique-dp,Okawa08,Gielen17obli-dp,Murphy2015}. Here, a deep pool refers to a liquid layer of depth $h$ greater than the drop diameter $D$, typically corresponding to a dimensionless pool depth $h'=h/D\geq4$. In contrast, many studies consider impact on thin liquid films with $h'<1$. For rainfall-driven deep-pool impacts, two issues are particularly important: how impact obliquity partitions the inertia that selects the macroscopic splashing regime, and how the breakup of crown rims and ligaments determines the size and population of secondary droplets.

The relative strengths of inertia, surface tension, and gravity are characterized by the Weber number $We=\rho D U^2/\sigma$ and the Froude number $Fr=U^2/(gD)$, where $\rho$ is the drop density, $U$ is the impact velocity, $\sigma$ is the surface tension of the receiving liquid, and $g$ is the gravitational acceleration. These parameters delineate distinct outcomes of normal impact \cite{Rein96,Hsiao88,Cresswell95,Liow01,Murphy2015}. Gentle impacts may result in floating or bouncing before coalescence \cite{Rein93rev,Rein96}. Following coalescence, a vortex ring can form \cite{Cresswell95}, whereas stronger impacts produce a cavity whose collapse can generate a central jet \cite{Hsiao88,Liow01} or entrain an air bubble \cite{Pumphrey90,Oguz90}. Together, these observations define the broad set of normal-impact regimes but do not account for the directional inertia introduced by oblique incidence.

At $We\gtrsim200$, rapid cavity expansion ejects a crown-shaped sheet whose rim develops fingers and ligaments that fragment into secondary droplets \cite{Liu2018JPO}. These droplets mediate air--sea exchange by transporting latent heat, carbon, and microplastics, and their evaporation can leave salt particles that act as cloud condensation nuclei \cite{Lehmann2021,Song2014,Trainic2020}. Crown-rim breakup must be distinguished from the thick Worthington jet generated by later cavity collapse: droplets shed from that central jet are relatively large and contribute little to air--sea fluxes \cite{Worthington1908,Edgerton1954,Rein93rev,Manzello02,Okawa2006}. Accordingly, as in studies of thin-film impact \cite{Mundo95,Cossali1997,Yarin2006,Wang2010}, we identify splashing by the production of secondary droplets from the crown rim or its finger-like jets. Under realistic rainfall conditions, $We$ can exceed 2000 \cite{Liu2016JGR}, and high-$We$ impacts can also lead to bubble-canopy formation and the prompt ejection of fine droplets \cite{Murphy2015,Wang23dp-We2000,Stober25obliquefilm,Josserand2003}. Predicting such impacts therefore requires both a regime criterion that accounts for impact direction and a physical description of the resulting secondary-droplet statistics.

Oblique impact, whether on a deep pool \cite{Zhbankova90,Leneweit05oblique-dp,Okawa08,Ray2012,Gielen17obli-dp,liu2018prf} or a thin film \cite{Chen20oblifilm, Stober25obliquefilm,Brambilla2013}, has received less attention than normal impact, despite its greater relevance to oceanic rainfall, in which surface winds impart horizontal momentum to falling drops. In oblique impact, the impact angle $\theta$, measured from the horizontal plane, is an additional parameter, and the $We$--$\theta$ plane has often been used to delineate different dynamic regimes \cite{Zhbankova90, Leneweit05oblique-dp, Okawa08, Gielen17obli-dp, Stober25obliquefilm}. At low Weber numbers ($We<2$), rebound and coalescence regimes were identified for $\theta\in[0,80^\circ]$ \cite{Zhbankova90}. The spreading and immersion of the drop liquid into the receiving liquid were examined for $We\in[15,249]$ and $\theta\in[5^\circ,64^\circ]$ \cite{Leneweit05oblique-dp}. 
For splashing at higher $We$, Okawa \textit{et al.}~\cite{Okawa08} used the splash index $K=We Oh^{-2/5}$ to predict the generation of secondary droplets from the crown rim, building on earlier splash criteria \cite{Yarin2006,Cossali1997}. Here, $Oh=\mu/\sqrt{D\rho\sigma}$ is the Ohnesorge number and $\mu$ is the dynamic viscosity. When $K<2100$, the drop coalesces with the receiving liquid, a behavior termed deposition \cite{Okawa08,Gielen17obli-dp,Stober25obliquefilm}; otherwise, splashing occurs. Over a narrow range of $Oh$, $K$ varies primarily with $We$. However, laboratory experiments have demonstrated a distinct $\theta$ dependence of the critical $K$ for splashing \cite{Okawa08}.


Beyond the $K=2100$ threshold, splashing can be broadly classified as front splashing (FS) or crown splashing (CS) \cite{Gielen17obli-dp}. FS corresponds to droplet generation only downstream of the impact point, whereas CS describes droplet ejection around the impact point in all directions.
Gielen \textit{et al.} showed that both the deposition--FS limit and the FS--CS transition can be expressed by a scaling law of the form $We Re^{1/4}$ with $\theta$-dependent coefficients, where $Re=\rho UD/\mu$ is the Reynolds number \cite{Gielen17obli-dp}.
Notably, both the $We Oh^{-2/5}$ scaling of Okawa \textit{et al.} and the $We Re^{1/4}$ scaling of Gielen \textit{et al.} originate from deposition--splash criteria established for impact on a dry solid wall \cite{Stow1980, Mundo95, Josserand2016} and were later extended to liquid-film impact \cite{Josserand2003, Thoroddsen2002}.
In addition to the three regimes mentioned above, side splashing (SS) and side-front splashing (SFS) have been identified for film impacts \cite{Stober25obliquefilm}. SS refers to droplet generation only to either side of the impact point, whereas SFS describes droplet generation both downstream of the impact point and to either side of it.
Specifically, the deposition--FS limit $We\cos\theta=128$ and the FS--SFS limit $We_N=450$ were proposed and showed good agreement with experiments. Here, $We_N=\rho U_N^2D/\sigma$ is the normal Weber number based on $U_N=U\sin\theta$. These results provide useful empirical boundaries for thin-film impact, but their applicability to deep-pool impacts and the physical mechanisms underlying the corresponding regime transitions remain to be established.

Secondary-droplet production presents a separate unresolved question. For normal impact onto a plane water surface, Okawa \textit{et al.}~\cite{Okawa2006} measured $N_s\propto K^{9/5}(h')^{-0.3}$, where $N_s$ is the number of secondary droplets and $h'$ is the dimensionless pool depth. At fixed $h'$ and nearly constant $Oh$, this relation reduces to $N_s\propto We^{9/5}$. Although subsequent studies have quantified how the number of secondary droplets and their size distributions vary with impact conditions \cite{li2019ds,zhang2021ds}, the physical origin of the $9/5$ exponent has not been established.

The present study addresses these two questions through three-dimensional simulations using the open-source code FluidX3D \cite{FluidX3D}. First, we determine the splashing regimes for oblique impact onto a deep pool and provide physical interpretations of the deposition--FS and FS--SFS transitions. Second, we reproduce the experimentally observed $N_s\propto We^{9/5}$ relation and develop a scaling interpretation based on the Weber-number dependence of the median droplet size, capillary--inertial breakup, and a volume-conservation closure.
\Cref{sec2} presents the numerical model, impact configuration, and qualitative validation against the laboratory observations reported by Liu \cite{Liu2018JPO}. \Cref{sec3} classifies the post-impact outcomes, constructs the regime map, and derives physics-based criteria for the deposition--FS and FS--SFS transitions. \Cref{sec4} examines the Weber-number dependence of the secondary-droplet population. It quantifies the median droplet size and droplet number and interprets their scaling using longitudinal velocity structure functions and capillary--inertial breakup. The main conclusions are summarized in \cref{sec5}.

\begin{figure}
\centering
\includegraphics[width=0.9\textwidth]{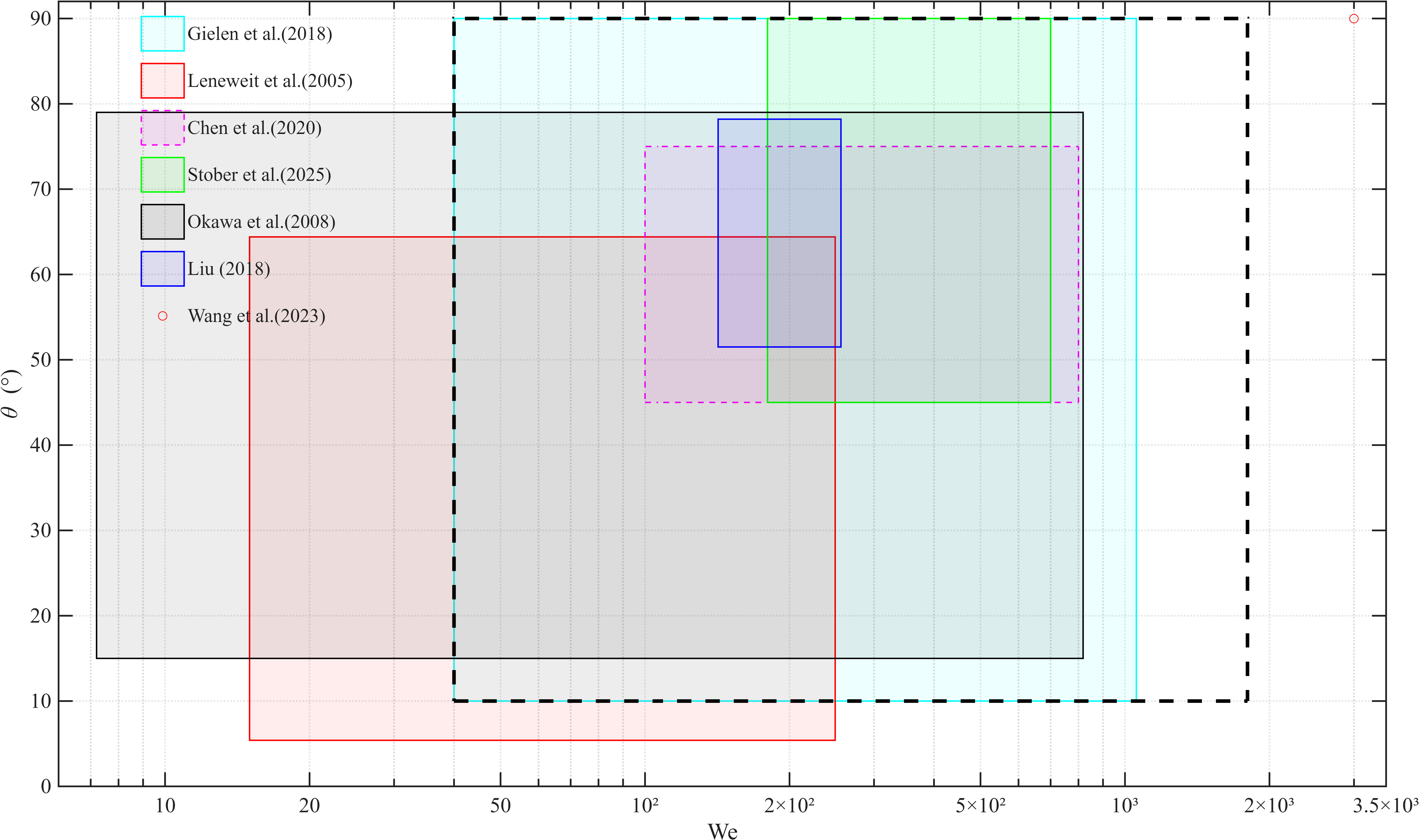}
\caption{Comparison of the parameter ranges in the $We$--$\theta$ plane explored in the present work (black box with a thick dashed outline) with those from previous studies. The dashed box represents the parameter range covered by the numerical simulations of Chen \textit{et al.}~\cite{Chen20oblifilm}, whereas the boxes with solid outlines represent the ranges covered by the experimental studies \cite{Leneweit05oblique-dp,Liu2018JPO,Stober25obliquefilm,Gielen17obli-dp,Okawa08}. The red dot in the upper-right corner indicates the high-$We$ impact experiment reported by Wang \textit{et al.}~\cite{Wang23dp-We2000}.
}
\label{Fig2}
\end{figure}

\section{Numerical methodology and model validation \label{sec2}}
\subsection{Numerical model and computational setup}
To simulate the oblique impact of a water drop onto a deep liquid pool, we used the GPU-accelerated solver FluidX3D, which implements a free-surface lattice Boltzmann method (LBM) with a volume-fraction-based interface treatment \cite{FluidX3D,lehmann2021hight,schwarzmeier2023comparison}. The LBM describes fluid motion through the evolution of particle distribution functions on a discrete velocity lattice, from which the macroscopic density and momentum are recovered \cite{Chen1998LBM,He1997LBM,krueger2017book}.
The collision process is modeled using the single-relaxation-time Bhatnagar--Gross--Krook 
(BGK) approximation \cite{Qian_1992BGK}. A D3Q27 lattice, consisting of 27 discrete 
velocity directions in three dimensions, is used to provide sufficient isotropy for the 
three-dimensional free-surface flow simulations \cite{krueger2017book,lehmann2021hight}. 
In the low-Mach-number limit, the LBM formulation recovers the incompressible 
Navier--Stokes equations through the Chapman--Enskog expansion 
\cite{He1997LBM,Chen1998LBM,krueger2017book}.
Details of the free-surface LBM equations can be found in the literature \cite{Qian_1992BGK,He1997LBM,Chen1998LBM,krueger2017book,lehmann2021hight,schwarzmeier2023comparison}.
The computational domain is divided into liquid, interface, and gas regions. The LBM solver is applied to the liquid phase, while the gas phase is treated as dynamically passive. FluidX3D tracks the liquid--gas interface using a sharp-interface volume-of-fluid scheme with a single-cell-thick interfacial layer and a local fill fraction $\phi$: liquid, interface, and gas cells satisfy $\phi=1$, $0<\phi<1$, and $\phi=0$, respectively \cite{FluidX3D,lehmann2021hight,schwarzmeier2023comparison}. Analytic piecewise-linear interface construction is used to reconstruct the interface for curvature evaluation \cite{Lehmann2022}, and surface tension is imposed through the Young--Laplace pressure jump $\Delta p=2\sigma\kappa$, where $\sigma$ is the surface tension coefficient and $\kappa$ is the local interface curvature \cite{lehmann2021hight}. FluidX3D has been applied to high-resolution simulations of raindrop-induced ejection of marine microplastics \cite{Lehmann2021}; its free-surface and curvature algorithms have also been assessed in dedicated benchmarking and methodological studies \cite{lehmann2021hight,schwarzmeier2023comparison,Lehmann2022}.

\begin{table}[htbp]
\caption{Material properties and impact conditions used in the simulations.}
\label{tab:parameters}
\begin{ruledtabular}
\begin{tabular}{lcc}
\hline
Parameter & Symbol & Value / Range \\
\hline
Impact velocity & $U$ & $1$--$8~\mathrm{m/s}$\\
Initial drop diameter & $D$ & $2.0$--$4.1~\mathrm{mm}$ \\
Impact angle & $\theta=\tan^{-1}(U_N/U_T)$ & $10^{\circ}$--$90^{\circ}$ \\
Froude number & $Fr=U^2/(gD)$ & $34$--$3262$ \\
Weber number & $We=\rho U^2D/\sigma$ & $41$--$1790$ \\
Normal Weber number & $We_N=\rho U_N^2D/\sigma$ & $5.8$--$1711$ \\
Tangential Weber number & $We_T=\rho U_T^2D/\sigma$ & $0$--$1573$ \\
Ohnesorge number & $Oh=\mu/(\rho\sigma D)^{1/2}$ & $0.0017$--$0.0027$ \\
\hline
\end{tabular}
\end{ruledtabular}
\footnotetext{Constant parameters: ambient temperature $T = 20\,^{\circ}\mathrm{C}$, water density $\rho = 1000~\mathrm{kg/m^{3}}$, air density $\rho_a = 1.204~\mathrm{kg/m^{3}}$, kinematic viscosity $\nu = 1.002\times10^{-6}~\mathrm{m^{2}/s}$, surface tension $\sigma = 7.3\times10^{-2}~\mathrm{N/m}$, and dimensionless pool depth $h'=h/D=4$. Here, $U_N=U\sin\theta$ and $U_T=U\cos\theta$ are the normal and tangential components of the impact velocity $U$, respectively.}
\end{table}


The computational domain is $10D \times 10D \times 8.5D$, with a prescribed dimensionless pool depth of $h'=h/D=4$. A no-slip bounce-back boundary condition is applied at the bottom wall, while periodic boundary conditions are imposed in the lateral directions. The simulations are organized into two data sets with different spatial resolutions. The coarse-resolution data set, computed with $\Delta x=20~\mu\mathrm{m}$, is used in \cref{sec3}, including construction of the regime map in \cref{Fig6:regime_map}. A separate set of 43 high-resolution cases is computed with $\Delta x=10~\mu\mathrm{m}$. Both data sets capture secondary droplets with diameters of approximately $50~\mu\mathrm{m}$, consistent with sizes observed in previous experiments \cite{Okawa08,Liu2018JPO,Rein96,Stober25obliquefilm,Marmottant2004}. The high-resolution grid additionally resolves the breakup of liquid jets into secondary droplets and is therefore used in \cref{sec4} for the secondary-droplet statistics.
The initial drop diameter is selected from the range $2.0\ \mathrm{mm}\leq D\leq4.1\ \mathrm{mm}$, based on reported raindrop size distributions \cite{Yakubu2016}. To account for aerodynamic deformation during free fall, drops are initialized as ellipsoids with an aspect ratio of 1.05~\cite{Wang23dp-We2000}. The Ohnesorge number falls within $0.0017\leq Oh\leq0.0027$. The impact velocity is varied over $1\ \mathrm{m/s}<U<8\ \mathrm{m/s}$, consistent with experimental measurements \cite{Liu2018JPO,Porc2013}. The parameter ranges explored in the present simulations largely cover those considered in earlier work (\cref{Fig2}). The material properties and impact conditions are summarized in \cref{tab:parameters}.

\subsection{Validation against laboratory observations and identification of splashing}

\begin{figure}[htbp]
\centering  
\includegraphics[width=0.6\textwidth]{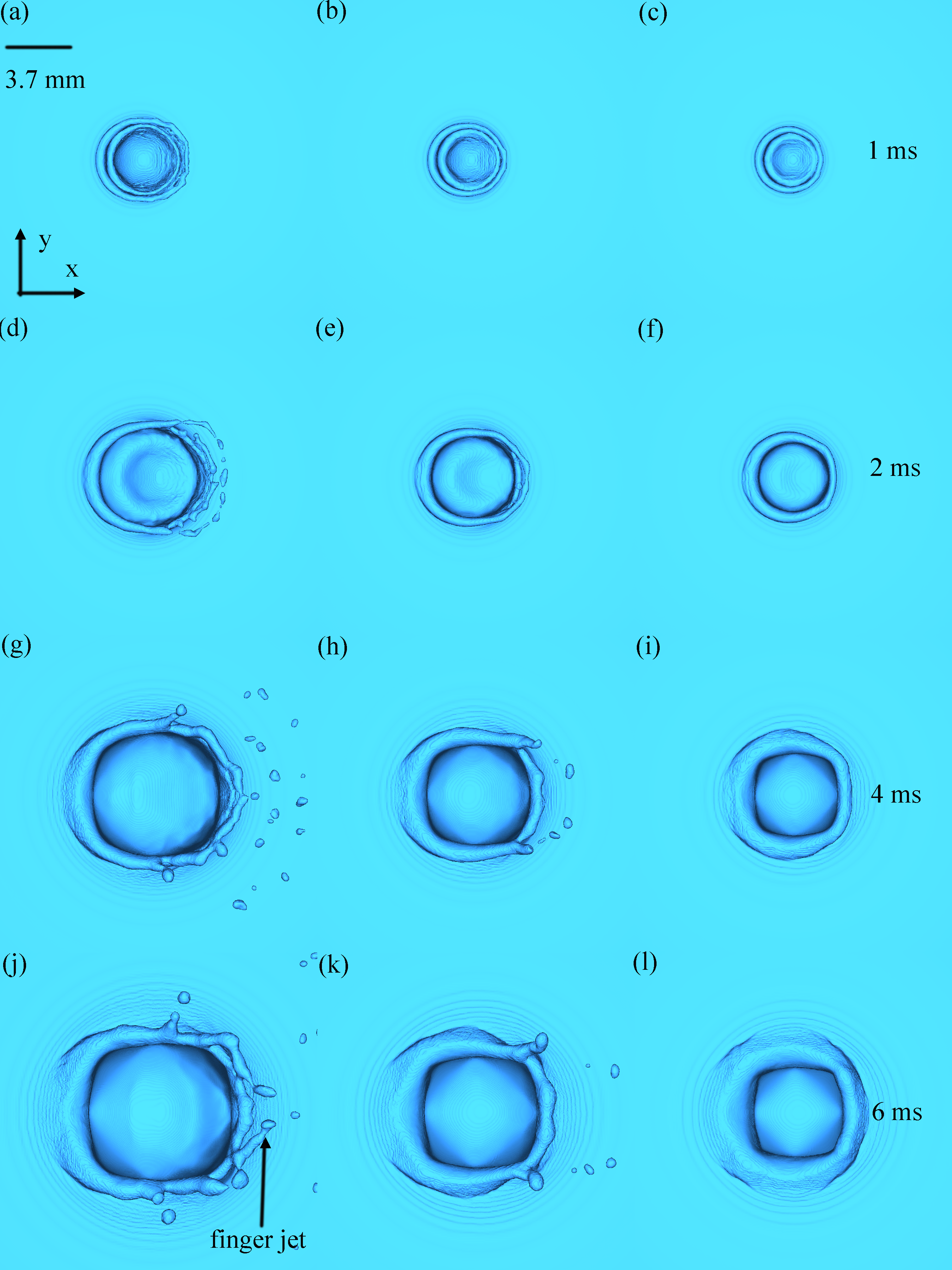} 
\caption{Numerical top-view sequences for three oblique drop-impact cases simulated using the impact kinematics measured by Liu \cite{Liu2018JPO}. From left to right, the columns correspond to ($D=3.7~\mathrm{mm}$, $\theta=73.3^\circ$, $U_T=0.58~\mathrm{m/s}$), ($D=3.2~\mathrm{mm}$, $\theta=70.6^\circ$, $U_T=0.71~\mathrm{m/s}$), and ($D=2.6~\mathrm{mm}$, $\theta=66.2^\circ$, $U_T=0.83~\mathrm{m/s}$). From top to bottom, the rows correspond to $t=1$, $2$, $4$, and $6~\mathrm{ms}$. The positive $x$ direction is aligned with the tangential impact velocity. The experimental sequences used for qualitative comparison are presented in Liu's Fig.~3.}
  \label{Fig5:validation-liu} 
\end{figure}

The numerical model was assessed against laboratory observations of oblique drop impact onto a deep pool reported by Liu \cite{Liu2018JPO}. Three representative cases were simulated using the measured drop diameters, impact angles, and tangential impact velocities. The experimental sequences are presented in Liu's Fig.~3, and the corresponding numerical top-view images at $t=1$, $2$, $4$, and $6~\mathrm{ms}$ are shown in \cref{Fig5:validation-liu}. The qualitative comparison focuses on crown asymmetry, downstream rim breakup, and secondary-droplet generation during the first $6~\mathrm{ms}$ after impact. For the largest drop, $D=3.7~\mathrm{mm}$, the simulation reproduces the rapid formation of an asymmetric crown and the pronounced deformation of its downstream rim, as shown in panels (a), (d), (g), and (j) of \cref{Fig5:validation-liu}. Thin jet-like structures emerge by $t=2~\mathrm{ms}$ and continue to eject secondary droplets at $t=4$ and $6~\mathrm{ms}$. This sequence is consistent with the experimentally reported elongation and breakup of jets on the downstream side of the crown rim \cite{Liu2018JPO}. For the intermediate case, $D=3.2~\mathrm{mm}$, the simulation captures an asymmetric crown with weaker early jetting, followed by downstream and lateral rim breakup at later times; see panels (b), (e), (h), and (k) of \cref{Fig5:validation-liu}. Liu likewise observed a less vigorous splash than for the $D=3.7~\mathrm{mm}$ case, with fewer secondary droplets generated from the downstream rim \cite{Liu2018JPO}. For the smallest drop, $D=2.6~\mathrm{mm}$, the simulated crown remains comparatively smooth and produces no resolved detached droplets during the interval shown in panels (c), (f), (i), and (l) of \cref{Fig5:validation-liu}; this weak-breakup morphology is similar to that observed experimentally \cite{Liu2018JPO}. Overall, the simulations reproduce the experimentally observed crown asymmetry and the systematic weakening of secondary-droplet production as $D$ decreases. This qualitative agreement supports the use of the model to construct the regime map in the $We$--$\theta$ parameter space. Here, splashing is identified by detached secondary droplets generated from the crown rim or its finger-like jets, excluding droplets produced by later cavity collapse or a thick Worthington jet.

\section{Splashing regimes and transition criteria  \label{sec3}}

\subsection{Morphological classification and regime map}

\begin{figure}[htbp]
\centering
\includegraphics[width=0.9\textwidth]{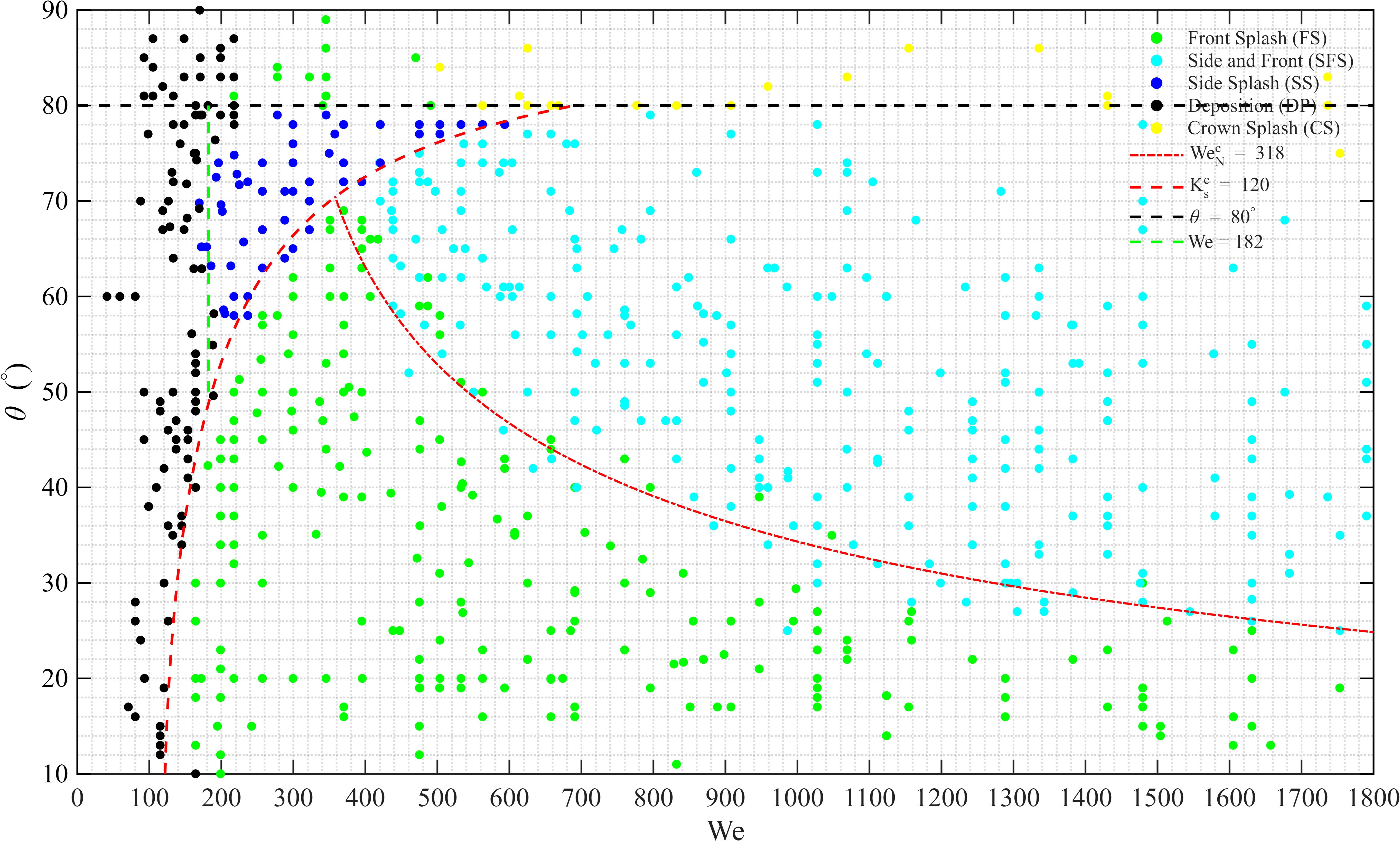}
\caption{Regime map in the $We$--$\theta$ parameter space showing five dynamic regimes: deposition, front splashing (FS), side splashing (SS), side-front splashing (SFS), and crown splashing (CS). These regimes are denoted by black, green, dark blue, light blue, and yellow dots, respectively. The red dashed line represents the deposition--FS limit given by $K_s^c=We\cos\theta=120$. The red dash-dotted line indicates the FS--SFS limit $We_N^c=318$.}
\label{Fig6:regime_map}
\end{figure}

In the parameter plane spanned by $We$ and $\theta$, Stober \textit{et al.} classified the post-impact dynamics ($\tau>4$) of an oblique drop impacting a liquid film into five regimes~\cite{Stober25obliquefilm}: deposition, front splashing (FS), side splashing (SS), side-front splashing (SFS), and crown splashing (CS). Deposition occurs at low $We$ when no secondary droplets are generated, whereas the other four regimes involve secondary-droplet generation. FS refers to secondary-droplet generation only downstream of the impact point (\cref{Fig8:splash}(a) and (e)). In the SS regime, droplets are generated on the lateral sides of the impact point. SFS occurs when secondary droplets are generated both downstream and on the lateral sides of the impact point. In the CS regime, secondary droplets are observed around the impact point in all directions.
Here, we adopt the same categorization and observe all five regimes in a series of simulations of deep-pool impacts. The explored Weber-number range, $41\leq We\leq1790$, and impact-angle range, $10^\circ\leq\theta\leq90^\circ$ (\cref{Fig6:regime_map}), are wider than those examined in the film-impact experiments \cite{Stober25obliquefilm}.

\begin{figure}[htbp]
\centering
\includegraphics[width=0.9\textwidth]{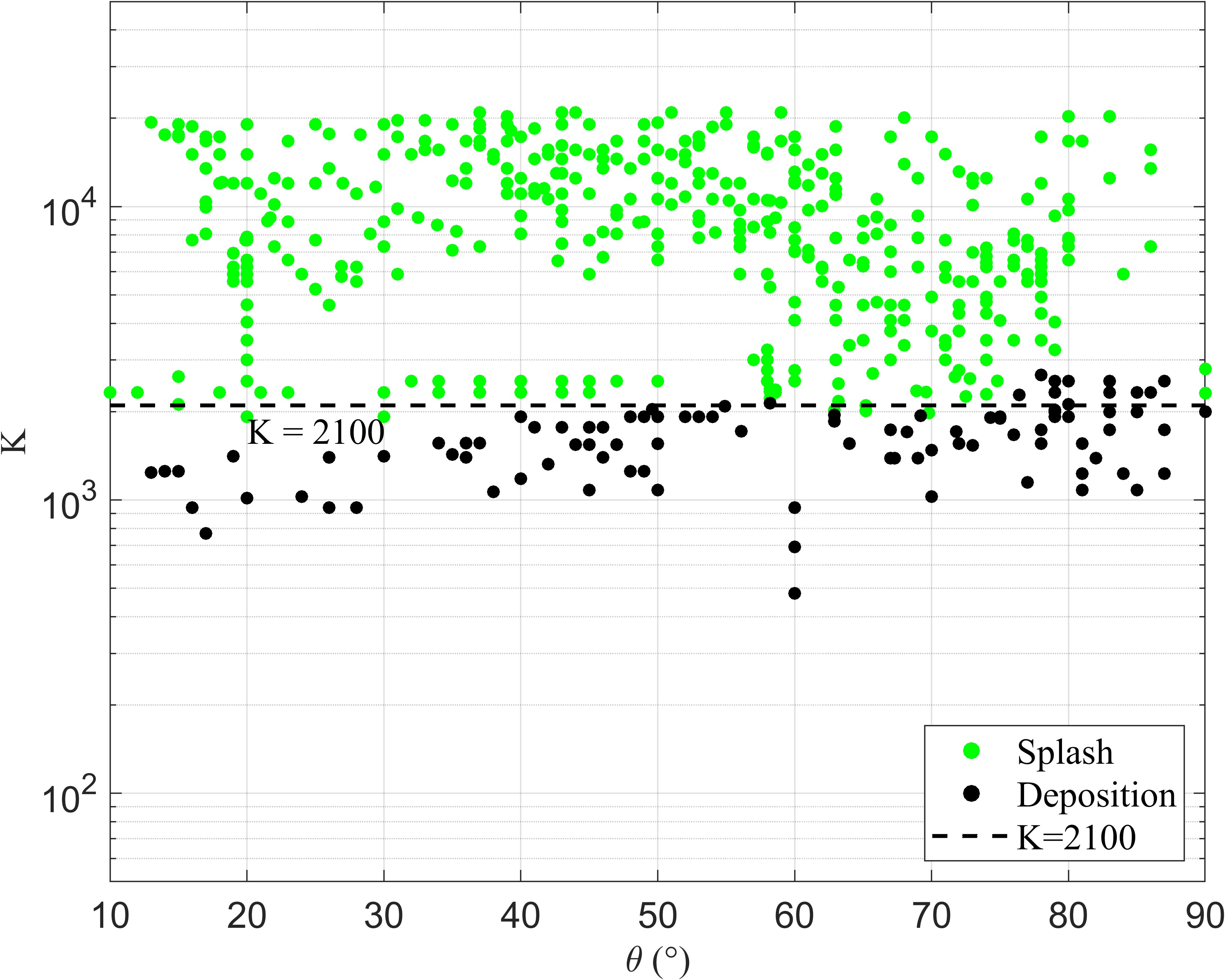}
\caption{Regime map in the $K$--$\theta$ plane. 
Black circles denote deposition, and green circles denote splashing. 
The dashed horizontal line marks the critical value 
$K_c = 2100$ proposed by Okawa \textit{et al.}~\cite{Okawa08}.}
\label{Fig7:K-$theta$}
\end{figure}

\begin{figure}[htbp]
\centering
\includegraphics[width=0.9\textwidth]{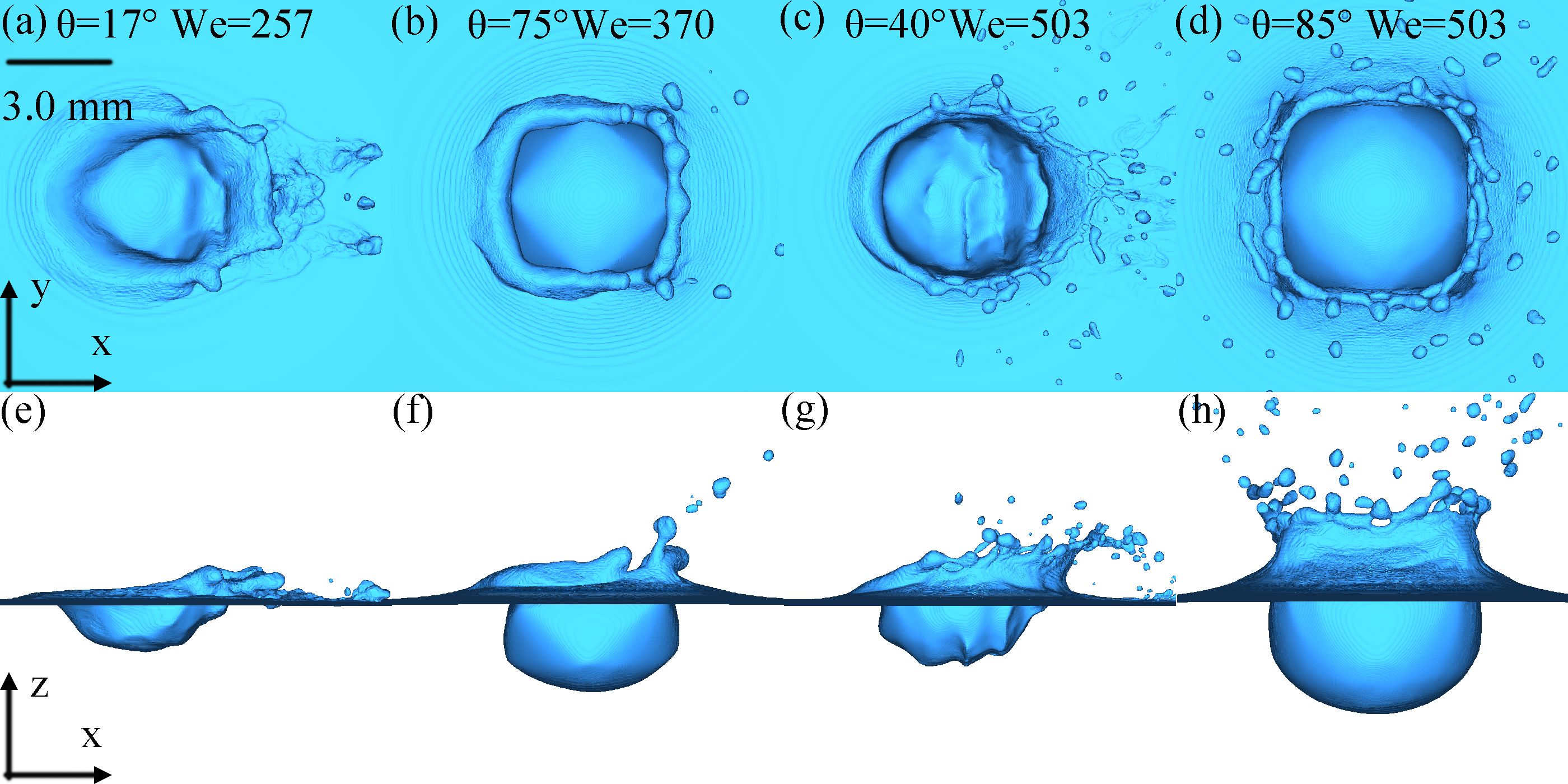}
\caption{Impacts of $D=3~\mathrm{mm}$ drops on a deep pool at different impact angles $\theta$ show distinct splashing morphologies. Panels (a)--(d) present the top views of front splashing (FS), side splashing (SS), side-front splashing (SFS), and crown splashing (CS), respectively. Panels (e)--(h) show the corresponding side views.}
\label{Fig8:splash}
\end{figure}

Previous experiments on deep-pool impacts \cite{Rein96,Rein93rev,Hallett84,Liu2018JPO} have identified a critical Weber number $We_c\approx182$ for the onset of splashing. Our simulation results agree with these studies, although a higher threshold, $We_c=400$, has also been reported \cite{Gielen17obli-dp}.
Instead of $We_c$, Okawa \textit{et al.}~\cite{Okawa2006,Okawa08} proposed an impact parameter $K$,
\begin{equation} \label{eq K_c}
K = We Oh^{-2/5} ,
\end{equation}
whose critical value, $K_c=2100$, delineates the deposition--splashing limit: splashing occurs at $K>K_c$. 
Since the Ohnesorge number varies little across our simulations ($0.0017\leq Oh\leq0.0027$), substituting the average value $\bar{Oh}=0.0022$ into \cref{eq K_c} shows that $K_c=2100$ corresponds approximately to $We_c=182$. The green dashed line at $We_c\approx182$ in \cref{Fig6:regime_map} therefore represents the threshold $K_c=2100$ shown in \cref{Fig7:K-$theta$}. This threshold distinguishes the splashing regimes from the deposition regime for $30^{\circ} \le \theta \le 75^{\circ}$. Beyond this $\theta$ range, two deviations are evident. 
First, at $\theta \le 30^{\circ}$, splashing still occurs at $K<K_c$. In contrast, Okawa \textit{et al.}~\cite{Okawa08} reported mostly deposition at these low $\theta$ values. Their experimental imaging system was unable to resolve secondary droplets smaller than approximately $50~\mu\mathrm{m}$. Therefore, splashing events involving such small droplets may have gone undetected and thus been classified as deposition in their experiments.
Second, at $\theta\geq75^{\circ}$, deposition is observed at $K>K_c$. Deposition at $K=3700$ was reported in the film-impact experiment \cite{Stober25obliquefilm} for $\theta=90^\circ$ and $We=318$.
These deviations imply that a constant $K_c$, or equivalently a constant $We_c$, provides only a rough estimate of the deposition--splashing limit; the $\theta$ dependence of this limit should also be considered.

To account for the effect of $\theta$, Stober \textit{et al.}~\cite{Stober25obliquefilm} introduced
\begin{equation}
K_s = We\cos\theta 
\label{eq:Ks_def}
\end{equation}
 and proposed a critical value $K_s^c=128$ as the deposition--FS limit for film impact. Here, we use \cref{eq:Ks_def} as the fitting function for the data in \cref{Fig6:regime_map}. A smaller critical value
\begin{equation}
    K_s^c = 120
\label{eq:Ks_pool}
\end{equation}
is identified as the threshold separating the deposition and FS regimes for deep-pool impact. A detailed derivation of \cref{eq:Ks_def,eq:Ks_pool} is presented in \cref{subsec-limits}.
The coarser optical sampling in the experiments of Stober \textit{et al.}~\cite{Stober25obliquefilm} may have failed to capture some of the small droplets resolved in the present simulations, which could partly explain why the reported value $K_s^c=128$ is higher than $K_s^c=120$ in \cref{eq:Ks_pool}. In addition, our high-resolution simulations show that \cref{eq:Ks_pool} distinguishes FS from deposition for cases with $We<We_c=182$ and $10^\circ\leq\theta\leq53^\circ$, in parameter ranges below those examined experimentally ($210<We<450$ and $53^\circ<\theta<73^\circ$).

Within the parameter domain examined by Stober \textit{et al.}, \cref{eq:Ks_pool} separates the SS regime from the FS regime (\cref{Fig6:regime_map}). SS occurs at $K_s\leq K_s^c$ and higher $We_N$ (with $\theta\geq50^\circ$), whereas FS occurs at $K_s>K_s^c$ and lower $We_N$ (with $\theta<50^\circ$). The FS regime occupies most of the $We$--$\theta$ space. Its dominance is consistent with the experiments of Gielen \textit{et al.}~\cite{Gielen17obli-dp}.

Previous studies have shown that $We_N$ plays a key role in governing splashing behavior \cite{Brambilla2013,Rein96,Rein93rev}. A critical normal Weber number, $We_N^c=425$, was identified as the threshold separating SS from deposition in a film-impact experiment \cite{Stober25obliquefilm}, although a smaller value, $We_N^c=127$, was reported by Liang \textit{et al.}~\cite{Liang2013}, who used glycerol--water mixtures as working fluids.
In our simulations, the SS regime is observed at $We_N\approx 144$ ($We\approx 200$, $\theta\approx 58^\circ$). In the SS regime, the finger-like jets develop from the lateral sides of the crown rim and are arranged symmetrically about the downstream $x$ axis (\cref{Fig5:validation-liu}(j)). These jets subsequently destabilize and fragment into secondary droplets (\cref{Fig8:splash}(b) and (f)).

\begin{figure}[htbp]
\centering
\includegraphics[width=0.9\textwidth]{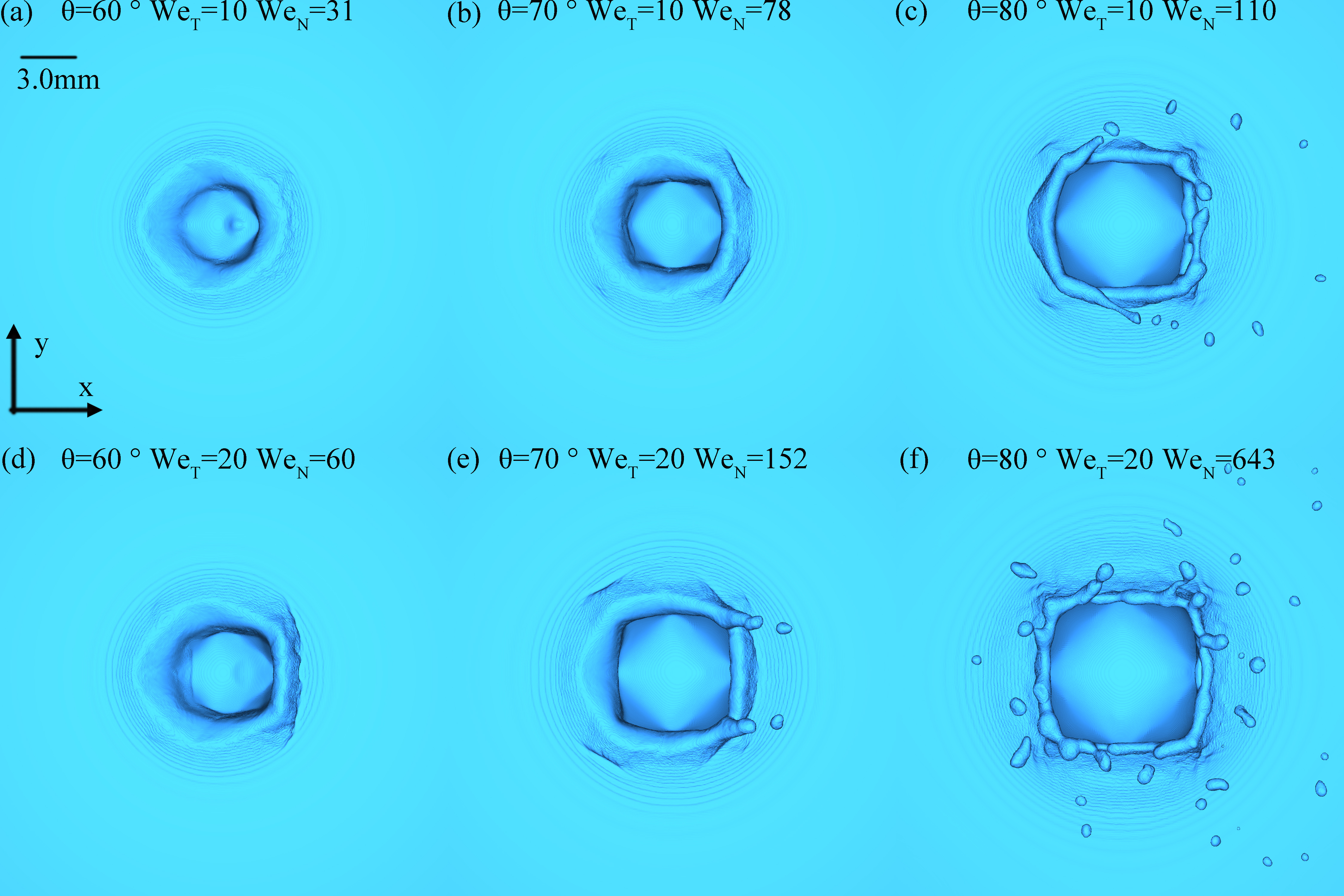}
\caption{Impacts of $D=3~\mathrm{mm}$ drops on a deep pool are shown for $\theta=60^\circ$ (left), $70^\circ$ (middle), and $80^\circ$ (right). The top panels (a)--(c) and the bottom panels (d)--(f) correspond to $We_T=10$ and $We_T=20$, respectively. All panels are shown at the dimensionless time $\tau=tU/D=4$.}
\label{Fig9:$We_T$-$theta$}
\end{figure}

When FS occurs (\cref{Fig8:splash}(a)), increasing $\theta$, and therefore $We_N$, can also trigger SS. In that case, secondary droplets are generated both downstream and on the lateral sides of the impact point. The concurrence of side and front splashing gives rise to the SFS regime (\cref{Fig8:splash}(c) and (g)). The SFS regime is separated from the FS regime by the critical normal Weber number $We_N^c=318$ (\cref{Fig6:regime_map}), whose derivation is presented in \cref{subsec-limits}.
As $We_N$ increases at fixed $We_T$, the dynamic regime evolves from deposition to SS and finally to CS. \Cref{Fig9:$We_T$-$theta$} shows that deposition occurs at $We_N\leq60$ (\cref{Fig9:$We_T$-$theta$}(a) and (d)). As $We_N$ increases from 152 (\cref{Fig9:$We_T$-$theta$}(e)) to 643 (\cref{Fig9:$We_T$-$theta$}(f)), the SS regime evolves into the CS regime.

When $\theta>80^\circ$, a nearly vertical impact with $We\leq200$ leads to deposition. Increasing $We$ initiates FS (\cref{Fig8:splash}(a) and (e)), and a further increase to $We=500$ leads to CS (\cref{Fig8:splash}(d) and (h)).
This dependence on $We$ is consistent with the experiments by Gielen \textit{et al.}~\cite{Gielen17obli-dp}, who reported deposition for $We < 400$, FS for $400 \le We \le 500$, and CS for $We > 500$ at $\theta = 85^\circ$. 

\subsection{Deposition--FS and FS--SFS limits \label{subsec-limits}}
\begin{figure}[htbp]
\centering
\includegraphics[width=0.9\textwidth]{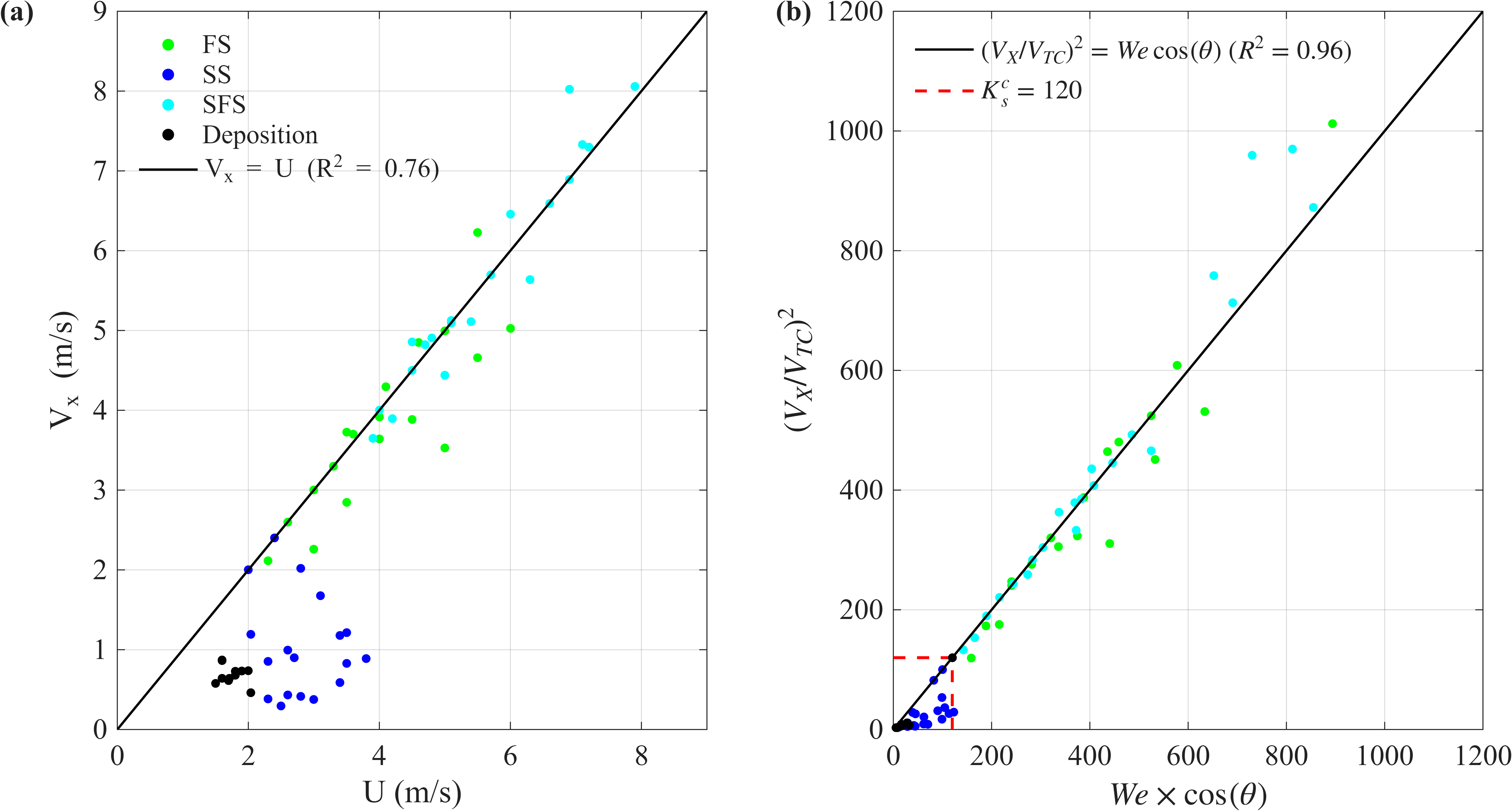}
\caption{(a) Linear relationship between the jet-tip velocity $V_x$ on the downstream side of the crown rim and the impact velocity $U$, with a coefficient of determination $R^2=0.76$. (b) Linear relationship between $V_x^2/V_{TC}^2$ and $K_s=We\cos\theta$, with $R^2=0.96$. The red dashed line denotes the critical value $K_s^c=V_x^2/V_{TC}^2=120$.}
\label{Fig10}
\end{figure}

We now derive the deposition--FS limit in \cref{eq:Ks_pool} based on the assumption made by Gielen \textit{et al.}~\cite{Gielen17obli-dp} that the crown-sheet volume flux $eDV$ is proportional to the drop-volume flux $D^2U$, such that
\begin{equation}
D^2 U \sim e D V ,
\label{eq:mass_balance}
\end{equation}
where $e$ is the crown-sheet thickness, which scales as
\begin{equation}
e \sim \sqrt{\frac{\nu D}{U}},
\end{equation}
and $V$ denotes the tip velocity of the finger-like jets that develop on the downstream side of the crown rim. Splashing occurs when the jet-tip velocity exceeds the Taylor--Culick velocity
\begin{equation}
V_{TC} \sim \sqrt{\frac{\sigma}{\rho e}},
\label{eq TC V}
\end{equation}
such that the threshold condition for FS can be written as \cite{Gielen17obli-dp,Yarin2006,Mundo95}
\begin{equation}
\frac{V}{V_{TC}} \sim We^{1/2} Re^{1/4}= We^{5/8} Oh^{-1/4} > K_G.
\end{equation}
Substituting the average Ohnesorge number $\bar{Oh}=0.0022$ and the critical Weber number $We_c=182$ for the onset of FS (\cref{Fig6:regime_map}) into the relation above gives $K_G\approx122$. This value is slightly lower than $K_G\approx130$, as reported by Gielen \textit{et al.}~\cite{Gielen17obli-dp} for normal impact.
For oblique impacts, the crown is asymmetric. The jet-tip velocity on the downstream side of the rim, $V_x$, exceeds that on the lateral sides, $V_y$~\cite{Brambilla2013,Gielen17obli-dp}. In such cases, the volume-flux relation in \cref{eq:mass_balance} can be projected onto the downstream ($x$) direction, yielding
\begin{equation}
D^2 U \cos\theta \sim e D V_x .
\label{eq:mass_projection}
\end{equation}
Substituting the expression for $e$ from \cref{eq:mass_projection} into \cref{eq TC V} and squaring both sides gives
\begin{equation}
\left( \frac{V_x}{V_{TC}} \right)^2 \sim \frac{\rho D U \cos\theta}{\sigma} V_x .
\label{eq:velocity_ratio_sq}
\end{equation}
Since $V_x\sim U$, as shown in \cref{Fig10}(a), the relation above becomes
\begin{equation}
\left( \frac{V_x}{V_{TC}} \right)^2 \sim We\cos\theta .
\label{eq:scaling_result}
\end{equation}
The right-hand side is $K_s$, as defined in \cref{eq:Ks_def}. The proportionality between $V_x^2/V_{TC}^2$ and $K_s$ is confirmed in \cref{Fig10}(b), which also shows that the threshold $K_s^c=120$ in \cref{eq:Ks_pool} serves as the deposition--FS limit for oblique impacts on a deep pool.

The concurrence of front and side splashing leads to the SFS regime. Therefore, the threshold separating SFS from FS can be derived by combining the definition of $K_s$ in \cref{eq:Ks_def} and the FS criterion in \cref{eq:Ks_pool} with the SS criterion.
Side splashing occurs when finger-like jets that develop on the lateral sides of the crown rim break up and generate secondary droplets. The kinetic energy of these jets can be estimated as $eDV_y^2$ and originates from the normal component of the drop kinetic energy, $D^2U_N^2$. Thus,
\begin{equation}
   e D V_y^2 \sim D^2 U_N^2,
\end{equation}
from which the square of the jet-tip velocity can be estimated as
\begin{equation}
    V_y^2 \sim \frac{D U_N^2}{e}.
\end{equation}
Since SS is triggered when $V_y$ exceeds the Taylor--Culick velocity given in \cref{eq TC V}, the threshold condition can be written as
\begin{equation}
\frac{V_y^2}{V_{TC}^2} \sim We_N \geq We_N^c.
\label{eq:SS-treshold}
\end{equation}
Since the SFS regime can be considered a superposition of the FS and SS regimes, it occurs when both \cref{eq:SS-treshold} and \cref{eq:Ks_pool} are satisfied. Substituting \cref{eq:Ks_pool} into \cref{eq:SS-treshold} yields the following criterion for the SFS regime:
\begin{equation}
    We_N > K_s^c \frac{\sin^2\theta}{\cos\theta}.
\label{eq:SFS-treshold}
\end{equation}
The boundary between the FS and SFS regimes spans $58^\circ\leq\theta\leq71^\circ$ (\cref{Fig6:regime_map}). Because the right-hand side of \cref{eq:SFS-treshold} increases monotonically with $\theta$, evaluating it at the largest boundary angle, $\theta=71^\circ$, gives $We_N^c\approx318$, which is used here as the characteristic SFS threshold.

\section{Weber-number scaling of the secondary-droplet population \label{sec4}}

\begin{figure}[htbp]
\centering
\includegraphics[width=0.9\textwidth]{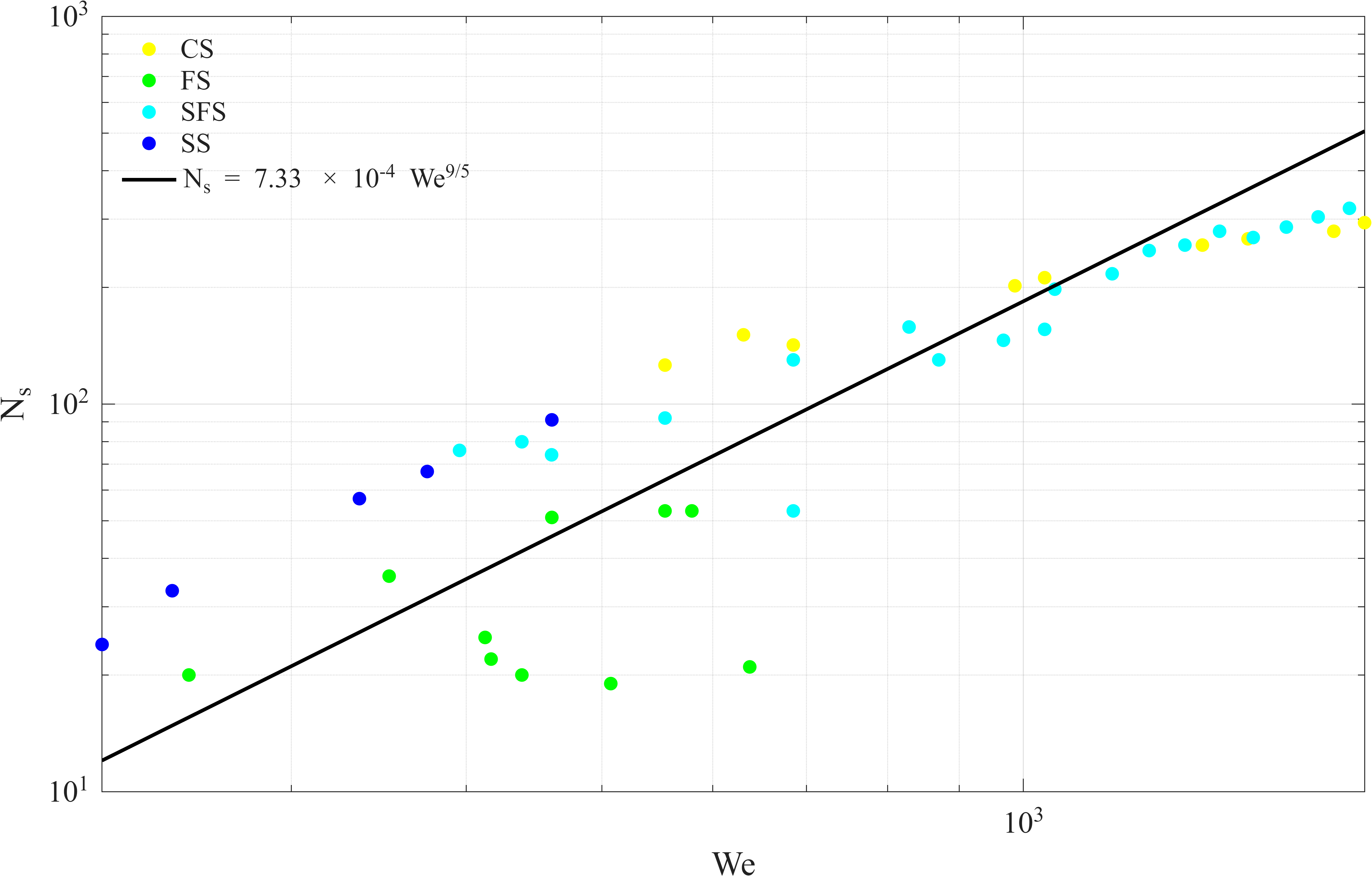}
\caption{Number of secondary droplets, \(N_s\), as a function of \(We\) for the FS, SFS, SS, and CS regimes. The solid line denotes \(N_s = 7.33\cdot 10^{-4}We^{9/5}\); the corresponding coefficient of determination is \(R^2 = 0.70\).}
\label{Fig11:Ns}
\end{figure}

The experiments by Okawa \textit{et al.}~\cite{Okawa2006} showed that the number of secondary droplets, $N_s$, increases systematically with impact intensity. For normal impacts, this dependence was described by the empirical relation
\begin{equation}
N_s = 7.84 \times 10^{-6} \cdot K^{9/5} \cdot (h')^{-0.3}.
\label{eq:Ns}
\end{equation}
Because $h'=4$ is fixed and $Oh$ varies only slightly across our simulations, variations in $K$ are governed primarily by $We$. Under these conditions, the relation above reduces to
\begin{equation}
N_s  \sim  We^{9/5}.
\label {eq Ns_We}
\end{equation}
Our simulations broadly reproduce this experimentally observed scaling across all four splashing regimes (\cref{Fig11:Ns}). All secondary-droplet statistics reported in this section are obtained from 43 high-resolution cases with $\Delta x=10~\mu\mathrm{m}$.

\begin{figure}[htbp]
\centering
\includegraphics[width=1.0\textwidth]{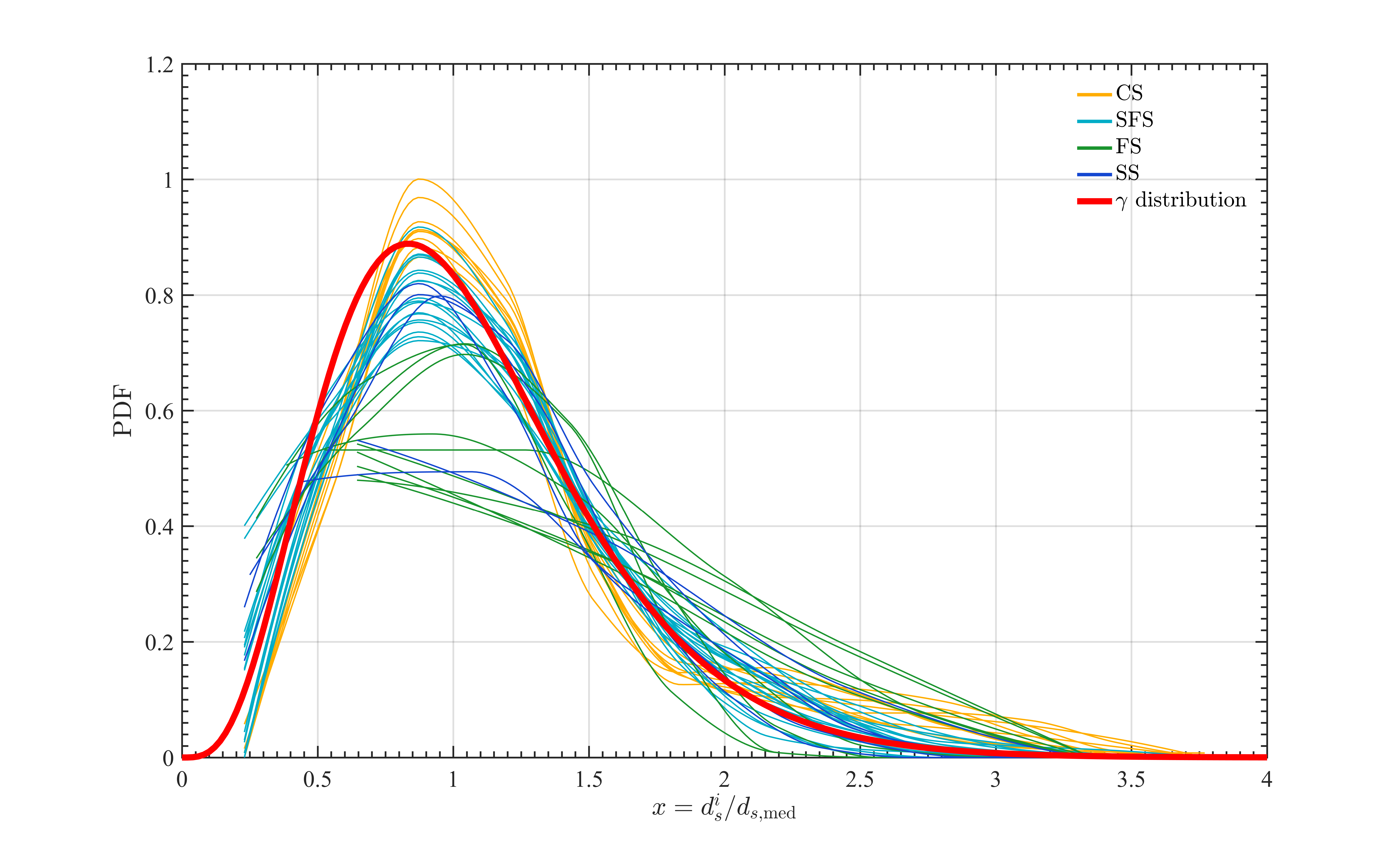}
\caption{Probability density functions (PDFs) of the secondary-droplet diameter normalized by the median, $x=d_s^i/d_{s,\mathrm{med}}$, for the individual impact cases in the CS, SFS, FS, and SS regimes. The red curve is a representative Gamma distribution, $\gamma(x;k,\beta)=x^{k-1}\exp(-x/\beta)/[\Gamma(k)\beta^k]$. Here, the shape parameter $k=4.6$ characterizes the breadth and skewness of this distribution, whereas the scale parameter $\beta=0.23$ sets the decay of its large-size tail.}
\label{Fig12:pdf}
\end{figure}

We next examine the characteristic secondary-droplet size and develop a scaling interpretation of the $We^{9/5}$ dependence based on the breakup dynamics. The size of the $i$th secondary droplet with volume $\mathcal{V}_i$ is quantified by its volume-equivalent spherical diameter,
\begin{equation}
d_s^i=\left(\frac{6\mathcal{V}_i}{\pi}\right)^{1/3}.
\end{equation}
To compare the distributions of $d_s^i$ among impact cases, a representative secondary-droplet diameter must be selected. The volume-averaged diameter $d_s^{\mathrm{va}}=\left(6\sum_{i=1}^{N_s}\mathcal{V}_i/\pi N_s\right)^{1/3}$ was used previously \cite{Okawa2006,Okawa08}. For a positively skewed population, however, the cubic weighting inherent in $d_s^{\mathrm{va}}$ makes this measure disproportionately sensitive to the sparse large-droplet tail, which has little influence on the droplet number $N_s$. The median diameter $d_{s,\mathrm{med}}$ is insensitive to these extreme sizes and is therefore selected as the typical secondary-droplet size. We define the normalized diameter as $x_i=d_s^i/d_{s,\mathrm{med}}$; its PDFs for the individual impacts are shown in \cref{Fig12:pdf}. Across the four splashing regimes, the distributions are consistently unimodal and positively skewed, with extended tails toward large diameters. Although their peak heights and widths vary among cases, they all exhibit a Gamma-like form rather than the approximately Gaussian distributions reported for liquid-film impacts \cite{li2019ds}. The red curve in \cref{Fig12:pdf} is included only as a representative Gamma distribution, not as evidence that all cases collapse onto a universal distribution.
The Gamma-like PDFs are consistent with ligament-mediated fragmentation \cite{villermaux2009single,Villermaux2004prl}. In this picture, strong variations in the local cross-sectional diameter produce a nonuniform distribution of liquid volume along corrugated rims and ligaments. Random rearrangement and aggregation of these liquid elements before capillary breakup generate a broad, positively skewed population of secondary droplets. The expanding sheets, irregular rims, and ligaments observed in the present deep-pool impacts constitute the fragmentation structures associated with this mechanism.

\begin{figure}[htbp]
\centering
\includegraphics[width=1.0\textwidth]{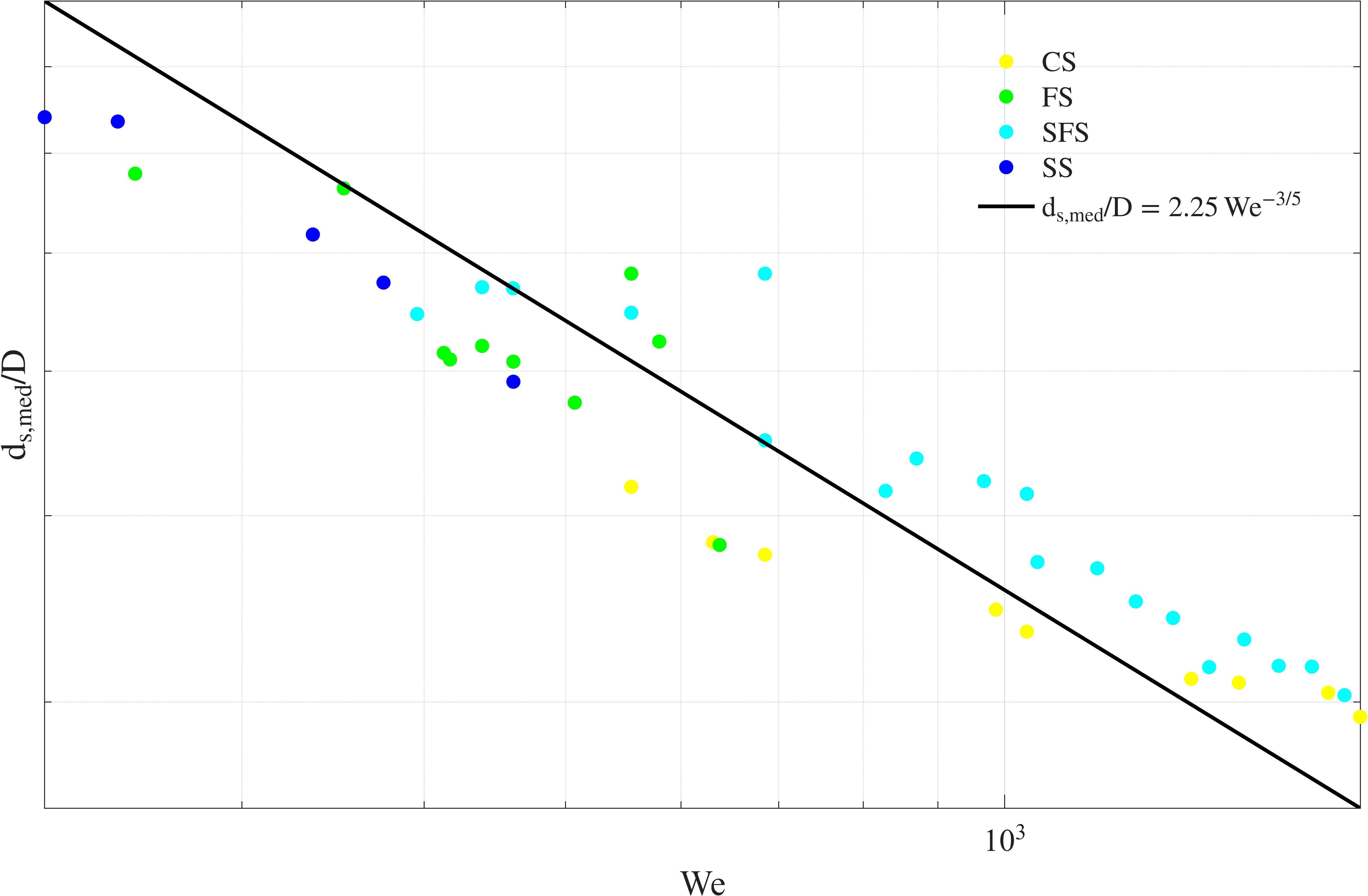}
\caption{Normalized median secondary-droplet diameter, \(d_{s,\mathrm{med}}/D\), as a function of \(We\) for the FS, SFS, SS, and CS regimes. The solid line denotes a fit with the exponent fixed at \(-3/5\); the corresponding coefficient of determination is \(R^2 = 0.76\).}
\label{Fig13:dsc_vs_We}
\end{figure}

\Cref{Fig13:dsc_vs_We} shows that $d_{s,\mathrm{med}}$ generally decreases with increasing $We$ across the splashing regimes, following
\begin{equation}
\frac{d_{s,\mathrm{med}}}{D}\sim We^{-3/5}.
\end{equation}
For solid-surface impacts, the dimensionless arithmetic-mean secondary-droplet diameter $d_{sa}$ was reported to scale as $d_{sa}\sim Re^{-1/2}$, a dependence attributed to the viscous length scale of the spreading lamella \cite{zhang2021ds}. $d_{sa}$ is distinct from both the volume-averaged diameter $d_s^{\mathrm{va}}$ defined above and the median diameter $d_{s,\mathrm{med}}$ used here. Substituting $Re=\sqrt{We}/Oh$ and $\bar{Oh}=0.0022$ transforms this relation into $d_{sa}\sim We^{-1/4}$. However, the observed $d_{s,\mathrm{med}}/D\sim We^{-3/5}$ dependence does not support the specific viscous-length scaling reported for solid-surface impacts. In a deep-pool impact, the ejecting sheet is not bounded by a no-slip solid surface. Instead, secondary droplets originate predominantly from freely deforming sheets, corrugated rims, and ligaments whose local thicknesses evolve through inertial stretching opposed by capillary restoration. Increasing $We$ strengthens inertial stretching and produces thinner structures, which are more prone to breakup into droplets. This mechanistic distinction motivates the capillary--inertial analysis below.

\begin{figure}[htbp]
\centering
\includegraphics[width=1.0\textwidth]{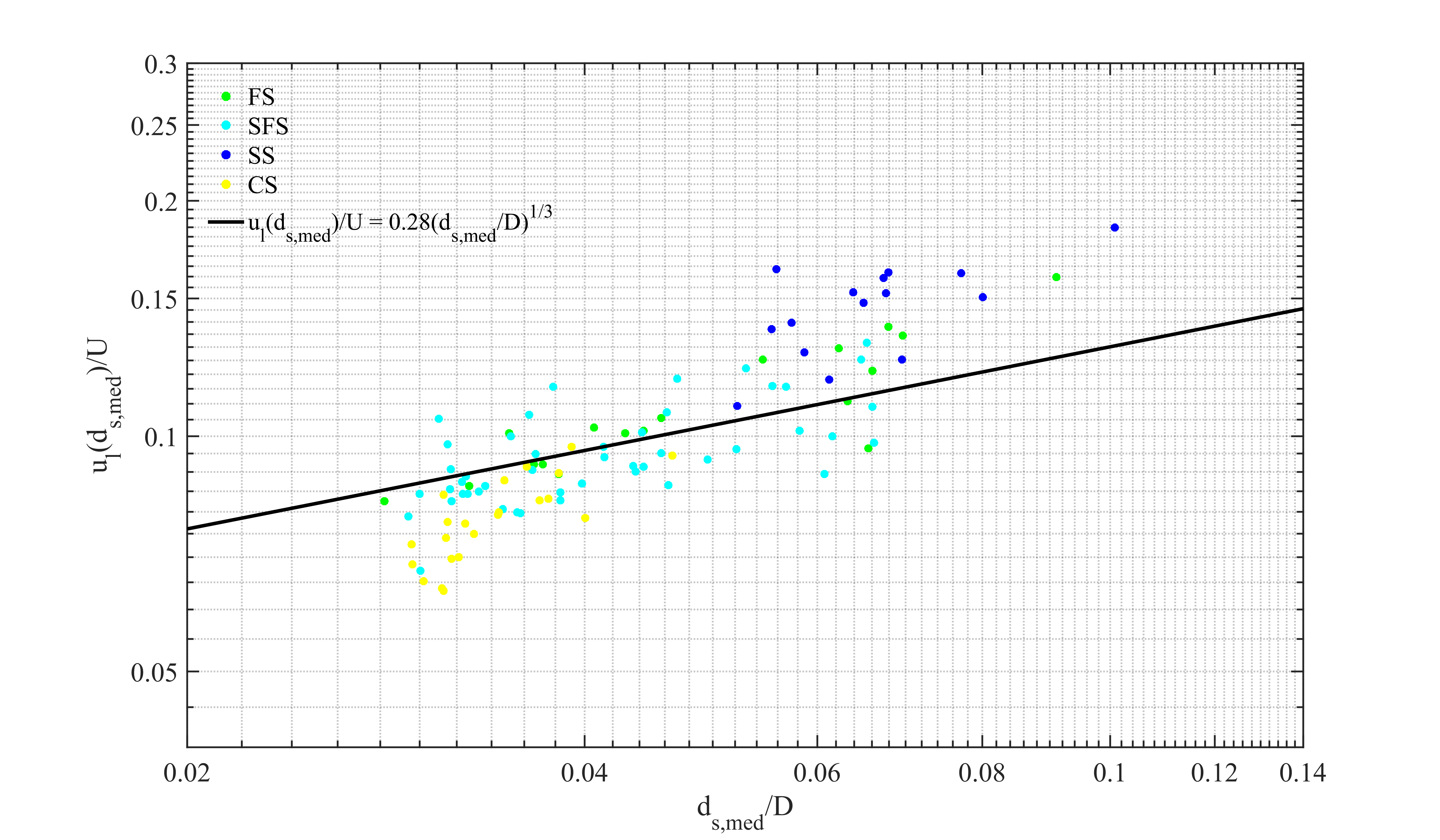}
\caption{Root-mean-square longitudinal velocity increment, $u_l(d_{s,\mathrm{med}})/U$, as a function of the dimensionless median secondary-droplet diameter, $d_{s,\mathrm{med}}/D$, for all individual structures in the FS, SFS, SS, and CS regimes. The solid line denotes $u_l(d_{s,\mathrm{med}})/U = 0.28(d_{s,\mathrm{med}}/D)^{1/3}$. The corresponding coefficient of determination is $R^2=0.55$.}
\label{Fig15:u_l_scaling}
\end{figure}

We next examine whether a capillary--inertial balance can account for the observed $We^{-3/5}$ scaling. At the characteristic breakup scale, we assume that the scale-dependent inertial force per unit length is locally balanced by the restoring surface tension,
\begin{equation}
\rho u_l^2(d_{s,\mathrm{med}})\,d_{s,\mathrm{med}}\sim\sigma,
\label{eq:capillary-inertial-breakup}
\end{equation}
where $u_l(d_{s,\mathrm{med}})$ denotes the root-mean-square longitudinal velocity increment over a separation of $d_{s,\mathrm{med}}$ and measures the local inertial fluctuations associated with breakup at that scale. To close \cref{eq:capillary-inertial-breakup}, $u_l^2(d_{s,\mathrm{med}})$ must be specified. Although the impact-induced flow is not fully developed turbulence, rapid sheet stretching and rim-finger formation generate velocity fluctuations over a broad range of scales. We therefore assume that $u_l(d_{s,\mathrm{med}})$ follows a one-third power-law scaling \cite{kolmogorov1991local, hinze1955fundamentals},
\begin{equation}
u_l(d_{s,\mathrm{med}})
\sim
U
\left(
\frac{d_{s,\mathrm{med}}}{D}
\right)^{1/3}.
\label{eq:one-third-scaling}
\end{equation}
To assess \cref{eq:one-third-scaling}, we calculated $u_l^2(d_{s,\mathrm{med}})$ using the second-order longitudinal velocity structure function
\begin{equation}
S_{2,L}^{(j)}(r)
=
\left\langle
\left[
\left(
\boldsymbol{u}(\boldsymbol{x}+\boldsymbol{r})
-
\boldsymbol{u}(\boldsymbol{x})
\right)
\cdot\hat{\boldsymbol{r}}
\right]^2
\right\rangle^{(j)},
\end{equation}
within each connected liquid structure responsible for secondary-droplet generation and located above the liquid surface. These structures were identified through three-dimensional connected-component labeling with a phase-field threshold of 0.5.
Here, superscript $j$ denotes the $j$th connected liquid structure.
$\boldsymbol{u}(\boldsymbol{x})$ is the velocity at position $\boldsymbol{x}$, $\boldsymbol{r}$ is the separation vector with magnitude $r$, $\hat{\boldsymbol{r}}=\boldsymbol{r}/r$ is the corresponding unit vector, and $\langle\cdot\rangle^{(j)}$ denotes an average over all valid pairs within the $j$th structure. To maintain adequate statistical sampling at different $r$, small structures containing fewer than 300 liquid voxels were excluded.
For the secondary droplets generated by the $j$th structure, the characteristic size is quantified by their median diameter $d_{s,\mathrm{med}}^{(j)}$. To associate $d_{s,\mathrm{med}}^{(j)}$ with the root-mean-square longitudinal velocity increment within the same parent structure, the velocity increment is written as
\begin{equation}
u_l^{(j)}
\left(
d_{s,\mathrm{med}}^{(j)}
\right)
=
\left[
S_{2,L}^{(j)}
\left(
d_{s,\mathrm{med}}^{(j)}
\right)
\right]^{1/2}.
\label{eq:ul_jet}
\end{equation}
Evaluating \cref{eq:ul_jet} for the individual structures yields a relation between $u_l$ and $d_{s,\mathrm{med}}$.
\Cref{Fig15:u_l_scaling} shows that the data obtained for liquid structures formed in the different splashing regimes are broadly consistent with the one-third scaling in \cref{eq:one-third-scaling}.

\begin{figure}[htbp]
\centering
\includegraphics[width=1.0\textwidth]{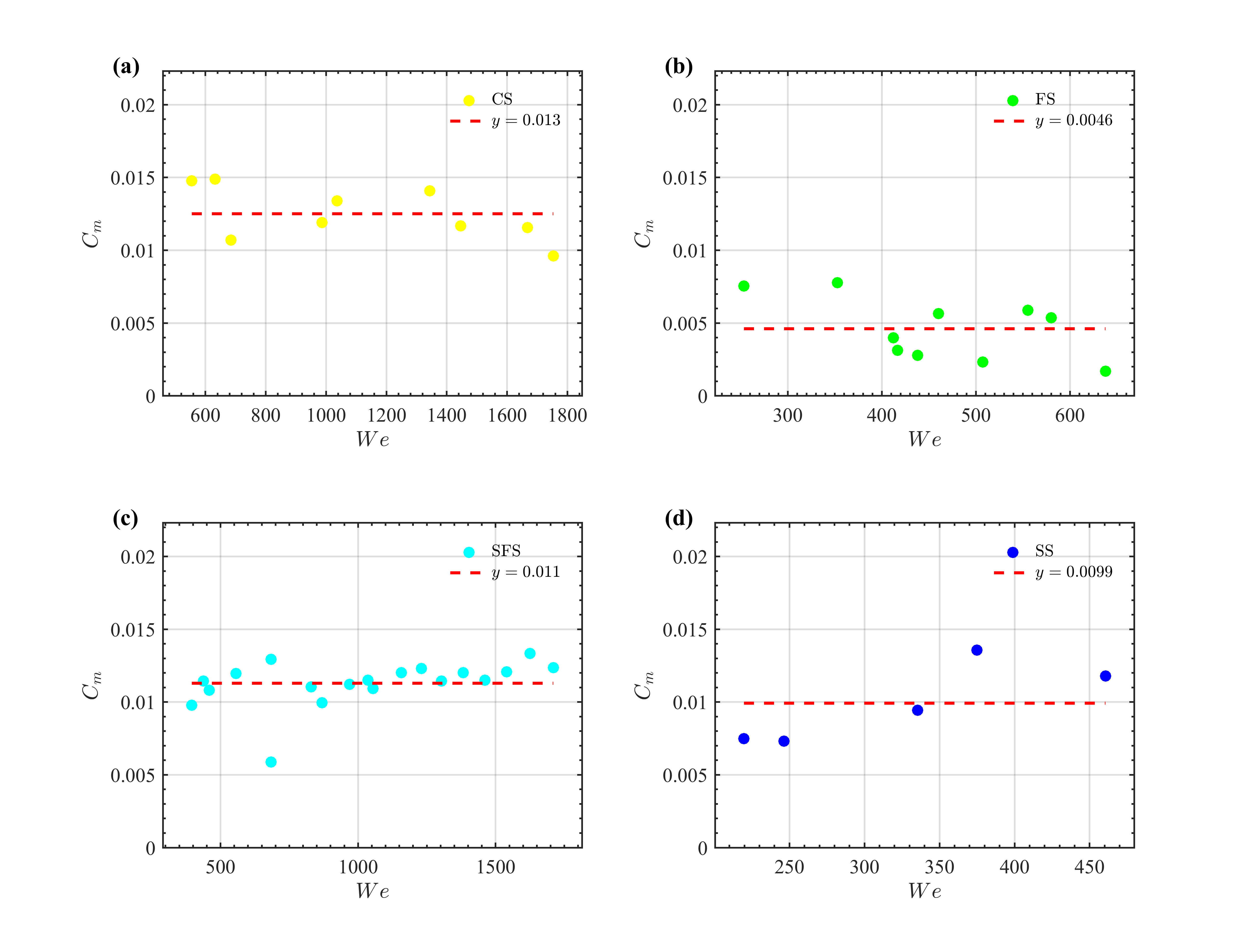}
\caption{The normalized total volume of the secondary droplets, $C_m=\sum_i\mathcal{V}_i/D^3$, as a function of $We$ for (a) CS, (b) FS, (c) SFS, and (d) SS. The red dashed lines denote the regime-specific mean values: $C_m=0.013$, $0.004$, $0.011$, and $0.010$, respectively.}
\label{Fig:secondary-volume-closure}
\end{figure}

Substituting \cref{eq:one-third-scaling} into the capillary--inertial balance in \cref{eq:capillary-inertial-breakup} yields
\begin{equation}
\rho U^2 D^{-2/3}d_{s,\mathrm{med}}^{5/3}\sim\sigma.
\end{equation}
Using $We=\rho U^2D/\sigma$, the relation above becomes
\begin{equation}
\frac{d_{s,\mathrm{med}}}{D}\sim We^{-3/5}.
\end{equation}
Data from the different splashing regimes generally collapse onto the predicted $-3/5$ scaling in \cref{Fig13:dsc_vs_We}. This collapse supports the proposed capillary--inertial interpretation of the typical droplet size across the different macroscopic morphologies.
To relate the droplet size to the droplet number, we close the scaling argument by assuming that the total mass of the secondary droplets remains proportional to the initial drop mass. Because the initial drop and the secondary droplets have the same density, this condition is equivalent to
\begin{equation}
\sum_{i=1}^{N_s}\mathcal{V}_i=C_mD^3,
\label{eq:secondary-volume-closure}
\end{equation}
where $C_m$ is the normalized total volume of the secondary droplets. \Cref{Fig:secondary-volume-closure} shows that $C_m$ is approximately independent of $We$ within each splashing regime. Although the proportion of the initial drop converted into secondary droplets differs among the regimes, its weak variation with $We$ supports treating $C_m$ as a regime-dependent constant for the present scaling analysis. Using $\mathcal{V}_i=(\pi/6)(d_s^i)^3$ and $x_i=d_s^i/d_{s,\mathrm{med}}$, the total secondary-droplet volume can be written exactly as
\begin{equation*}
\sum_{i=1}^{N_s}\mathcal{V}_i
=\frac{\pi}{6}N_s d_{s,\mathrm{med}}^3 M_3,
\qquad
M_3=\frac{1}{N_s}\sum_{i=1}^{N_s}x_i^3,
\end{equation*}
where $M_3$ is the dimensionless third moment of the normalized droplet-size distribution. The broadly similar normalized droplet-size distributions observed within each regime (\cref{Fig12:pdf}) suggest that $M_3$ varies only weakly with $We$. We therefore absorb $M_3$, together with the factor $\pi/6$, into a regime-dependent proportionality constant. Hence, $\sum_i\mathcal{V}_i\sim N_s d_{s,\mathrm{med}}^3$, and \cref{eq:secondary-volume-closure} gives
\begin{equation}
N_s \sim C_m\left(\frac{D}{d_{s,\mathrm{med}}}\right)^3.
\end{equation}
Substituting $d_{s,\mathrm{med}}/D\sim We^{-3/5}$ into this relation yields
\begin{equation}
N_s\sim We^{9/5},
\end{equation}
which recovers the $9/5$ scaling observed in previous experiments \cite{Okawa2006} and reproduced in the present simulations.

\clearpage
\section{Conclusions}\label{sec5}
We numerically investigated the oblique impact of water drops onto a deep quiescent pool over $41\leq We\leq1790$ and $10^\circ\leq\theta\leq90^\circ$, with a fixed dimensionless pool depth $h'=4$. The simulations reproduce the principal features observed experimentally, including asymmetric crown formation, downstream jetting, lateral fingering, and the onset of splashing. Five post-impact regimes are identified in the $We$--$\theta$ plane: deposition, front splashing (FS), side splashing (SS), side-front splashing (SFS), and crown splashing (CS). Their arrangement in the regime map shows that the total impact inertia alone is insufficient to characterize oblique impact; its tangential and normal components govern distinct parts of the crown response.

The deposition--FS transition is described by the tangential-inertial parameter $K_s=We\cos\theta$, with $K_s^c\approx120$ for the present deep-pool impacts. This threshold is close to that reported for liquid-film impacts \cite{Stober25obliquefilm}. The present analysis relates it to the competition between downstream crown-rim inertia and capillary retraction at the Taylor--Culick velocity. The onset of lateral splashing is instead controlled primarily by normal impact inertia. When the FS condition is also satisfied, a critical normal Weber number $We_N^c\approx318$ marks the FS--SFS transition, at which the front- and side-splashing modes coexist. These results show that the regime transitions can be interpreted through a directional partition of impact inertia: the tangential component promotes downstream ejection, whereas the normal component drives lateral fingering and the development of a more azimuthally distributed crown.

Beyond these regime-scale transitions, the simulations reveal fragmentation behavior shared across the four splashing regimes. The droplet-size distributions are positively skewed and are consistent with fragmentation of corrugated rims and ligaments. The median secondary-droplet diameter follows $d_{s,\mathrm{med}}/D\sim We^{-3/5}$. The second-order longitudinal velocity structure functions are broadly consistent with the scaling $u_l(d_{s,\mathrm{med}})/U\sim(d_{s,\mathrm{med}}/D)^{1/3}$ for the root-mean-square longitudinal velocity increment at the median-droplet scale. Combining this scale-dependent velocity with the capillary--inertial breakup condition recovers the observed $-3/5$ diameter scaling. Together, the approximately $We$-independent ratio of the total secondary-droplet mass to the initial drop mass and the assumed weak variation of the normalized size-distribution shape close the scaling argument and yield $N_s\sim We^{9/5}$, providing a scaling-based interpretation of the corresponding secondary-droplet-number dependence reported in previous experiments \cite{Okawa2006}.

The present results characterize two distinct levels of behavior in oblique deep-pool impact. At the scale of the crown, directional impact inertia determines where splashing first occurs and how the macroscopic regime changes with $We$ and $\theta$. At the scale of the rims and ligaments, the competition between velocity fluctuations and surface tension is consistent with the characteristic-size scaling, while the volume constraint links that size to the droplet population.

\begin{acknowledgments}
This work was supported by the National Natural Science Foundation of China (Grant No. 42276019) and the National Key Research and Development Program of China (Grant No. 2024YFC2817003).
\end{acknowledgments}

\section{Data Availability Statement}
The data that support the findings of this study are available from the corresponding author upon reasonable request.

\bibliography{aipsamp}

\end{document}